\documentclass[12pt]{article}
\usepackage[left= 2.5cm, right= 2.5cm, bottom=2.5cm, top=2.5cm]{geometry}
\usepackage{amsmath}
\usepackage{amssymb}
\usepackage{times}
\usepackage{amsfonts}
\usepackage{graphicx}
\usepackage{colordvi}
\usepackage[english]{babel}
\usepackage{verbatim}
\usepackage{color}
\usepackage{mathrsfs}
\usepackage{epsfig}
\usepackage{wrapfig}
\usepackage{floatflt}
\usepackage{enumitem}
\usepackage{setspace}
\usepackage{upgreek}
\usepackage{cite}
\usepackage[font=small,format=plain,justification=justified]{caption}
\let\oldhat\hat

\renewcommand{\hat}[1]{\oldhat{\boldsymbol{\mathbf{#1}}}}

\newcommand{\eref}[1]{Eq.~(\ref{eq:#1})}

\newcommand{\ux}{\mathbf{\hat{x}}}
\newcommand{\uy}{\mathbf{\hat{y}}}
\newcommand{\uz}{\mathbf{\hat{z}}}
\newcommand{\bfB}{\mathbf{B}}
\newcommand{\bfm}{\mathbf{m}}
\newcommand{\bfr}{\mathbf{r}}
\newcommand{\bfk}{\mathbf{k}}
\def\be{\begin{equation}}
\def\ee{\end{equation}}

\title{Imaging spin-wave damping underneath metals using electron spins in diamond} 
\vspace{12pt}

\author
{Iacopo Bertelli$^{1,2}$, Brecht G. Simon$^{1}$, Tao Yu$^{3}$, Jan Aarts$^{2}$, Gerrit E. W. Bauer$^{1,4}$,\\ Yaroslav M. Blanter$^{1,4}$ $\&$ Toeno van der Sar$^{1,*}$}
\begin{document} 
\baselineskip24pt
\maketitle
\noindent$^{1}$ Department of Quantum Nanoscience, Kavli Institute of Nanoscience, Delft University of Technology, Lorentzweg 1, 2628 CJ, Delft, The Netherlands\\
$^{2}$ Huygens - Kamerlingh Onnes Laboratorium, Leiden University, Niels Bohrweg 2, 2300 RA, Leiden, The Netherlands\\
$^{3}$ Max Planck Institute for the Structure and Dynamics of Matter, Luruper Chausee 149, 22761 Hamburg, Germany\\
$^{4}$  WPI-AIMR \& Institute for Materials Research \& CSRN, Tohoku University, Sendai 980-8577, Japan\\
$^*$ Corresponding author. Email:  T.vanderSar@tudelft.nl.
\onehalfspacing

\section*{Abstract}
Spin waves in magnetic insulators are low-damping signal carriers that could enable a new generation of spintronic devices. The excitation, control, and detection of spin waves by metal electrodes is crucial for interfacing these devices to electrical circuits. It is therefore important to understand metal-induced damping of spin-wave transport, but characterizing this process requires access to the underlying magnetic films. Here we show that spins in diamond enable imaging of spin waves that propagate underneath metals in magnetic insulators, and then use this capability to reveal a 100-fold increase in spin-wave damping. By analyzing spin-wave-induced currents in the metal, we derive an effective damping parameter that matches these observations well. We furthermore detect buried scattering centers, highlighting the technique's power for assessing spintronic device quality. Our results open new avenues for studying metal - spin-wave interaction and provide access to interfacial processes such as spin-wave injection via the spin-Hall effect.

\newpage

\section{Main Text}
\subsection*{Introduction}
\vspace{-6pt}
Spin waves are collective, wave-like excitations of the spins in magnetic materials\cite{Stancil}. The field of magnon spintronics aims at using these waves as signal carriers in information processing devices\cite{Chumak2015}. Since its recent inception, the field has matured rapidly\cite{Barman2021} and successfully realized prototypical spin-wave devices that implement logical operations\cite{Fischer2017,Talmelli2020,Wang2020,Chumak2014,Cornelissen2018}. In such devices, the spin waves are typically excited inductively\cite{Fischer2017,Talmelli2020,Wang2020,Chumak2014,Cornelissen2018} or via spin-pumping based on the spin-Hall effect\cite{Sinova2015,Cornelissen2015}, using electric currents in metal electrodes that are deposited on top of thin-film magnetic insulators. As such, it is a key challenge to understand the interaction between the metals and the spin waves in the magnetic insulators, but this requires the ability to study the buried magnetic films and is hampered by the opacity of the metals to optical probes.\\

\noindent We address this challenge using magnetic imaging based on electron spins in diamond\cite{Bertelli2020}. Metal films of sub-skin-depth thickness are transparent for microwave magnetic fields, which enables imaging of spin waves traveling underneath the metals by detecting their magnetic stray fields. We demonstrate this ability by imaging spin waves that travel underneath 200-nm-thick metal electrodes in a thin film of the magnetic insulator yttrium iron garnet (YIG). We find that the spatial spin-wave profiles under the metals reveal a surprisingly strong metal-induced spin-wave damping. By introducing the spin-wave-induced currents in the metal self-consistently into the Landau-Lifshitz-Gilbert (LLG) equation, we derive an analytical expression for the spin wave damping that matches our experimental observations without free parameters. We demonstrate that this eddy-current-induced damping mechanism dominates up to a threshold frequency above which three-magnon scattering becomes allowed and increases damping further. \\

\noindent Our imaging platform is an ensemble of shallowly implanted nitrogen-vacancy (NV) centers in diamond (Fig.~\ref{fig1-damping}a). NV centers are lattice defects with an $S=1$ electron spin that can be polarized by optical excitation, controlled by microwaves, and read out through spin-dependent photoluminescence\cite{Gruber1997,Rondin2014}. Since NV centers can exist within $\sim10$~nm from the surface of diamond\cite{Rosskopf2014}, they can be brought within close proximity to a material of interest. Combined with an excellent sensitivity to magnetic fields\cite{Rondin2014}, these properties make NV spins well suited for stray-field probing of spins and currents in condensed matter systems\cite{Casola2018}. 
\subsection*{Results}
\begin{figure}[h]
\begin{center}
{\includegraphics[width=1\textwidth]{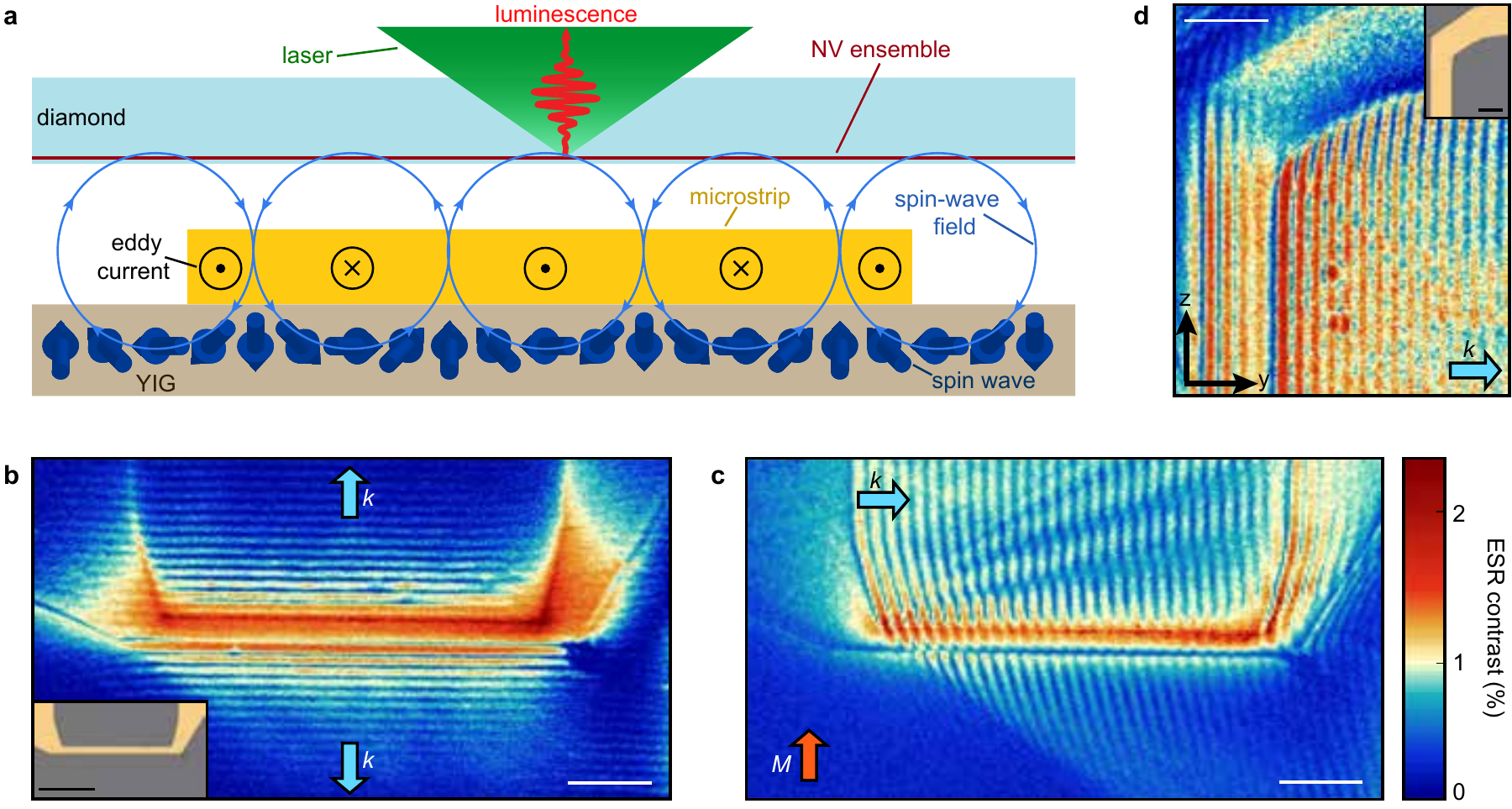}}
\end{center}
\caption{\textbf{Fig. 1. Magnetic imaging of microstrip excited spin waves using electron spins in diamond.} (\textbf{a}) Idea of the experiment. An ensemble of nitrogen-vacancy (NV) centers in a diamond chip is used to image the magnetic stray fields generated by spin waves in a YIG thin film. The ability to image spin waves underneath metals is used to study the metal-induced spin-wave damping. (\textbf{b})-(\textbf{d}) Spatial maps of the NV electron spin resonance (ESR) contrast when exciting spin waves resonant with an NV ESR transition. The oscillations result from the interference of the spin-wave and direct microstrip fields. The magnetization ($\mathbf{M}$) points along $z$. The directions of the predominantly excited spin-wave vectors ($\mathbf{k}$) are indicated. Scale bars: 20~$\upmu$m. (\textbf{b}) Backward volume waves ($\mathbf{k\parallel M}$), excited by applying a drive frequency $\omega/2\pi = 1.934$~GHz that is 0.17~GHz below the FMR at $B_0=33.5$~mT. Inset: micrograph of the sample (scale bar: 40~$\upmu$m). (\textbf{c}) Spin waves in the Damon-Eshbach configuration ($\mathbf{k \perp M}$) excited by applying a drive frequency $\omega/2\pi = 2.590$~GHz that is 1.12~GHz above the FMR at $B=15$~mT. (\textbf{d}) Spin waves underneath and next to a gold microstrip used for spin-wave excitation (inset). Scale bars: 20~$\upmu$m. As the skin depth of a 2.5~GHz magnetic field in gold is $\sim$1.6~$\upmu$m, spin waves are clearly visible underneath the 200~nm gold film.}
\label{fig1-damping}%
\end{figure}

To image propagating spin waves, we place a diamond membrane containing a layer of NV centers implanted $\sim10-20$~nm below the diamond surface onto a YIG film equipped with 200~nm thick gold microstrips (Methods). Passing a microwave current through a microstrip generates a magnetic field that excites spin waves in the YIG (Fig.~\ref{fig1-damping}a). These waves create a magnetic stray field that interferes with the direct microstrip field, leading to a spatial standing-wave pattern in the total amplitude of the oscillating magnetic field\cite{Bertelli2020}. We spatially map this amplitude by locally measuring the contrast of the NV electron spin resonance (ESR) transitions. By changing the drive frequency while adjusting the static magnetic field ($B_0$) to maintain resonance with the NV ESR frequency (Methods), we can excite and detect spin waves with wavevectors either along or perpendicular to the static magnetization $\mathrm{M}$ (Fig.~\ref{fig1-damping}b-c). The spin waves are clearly visible both underneath and next to the gold microstrips (Fig.~\ref{fig1-damping}b-d).\\

\begin{figure}[htbp]
\begin{center}
{\includegraphics[width=1\textwidth]{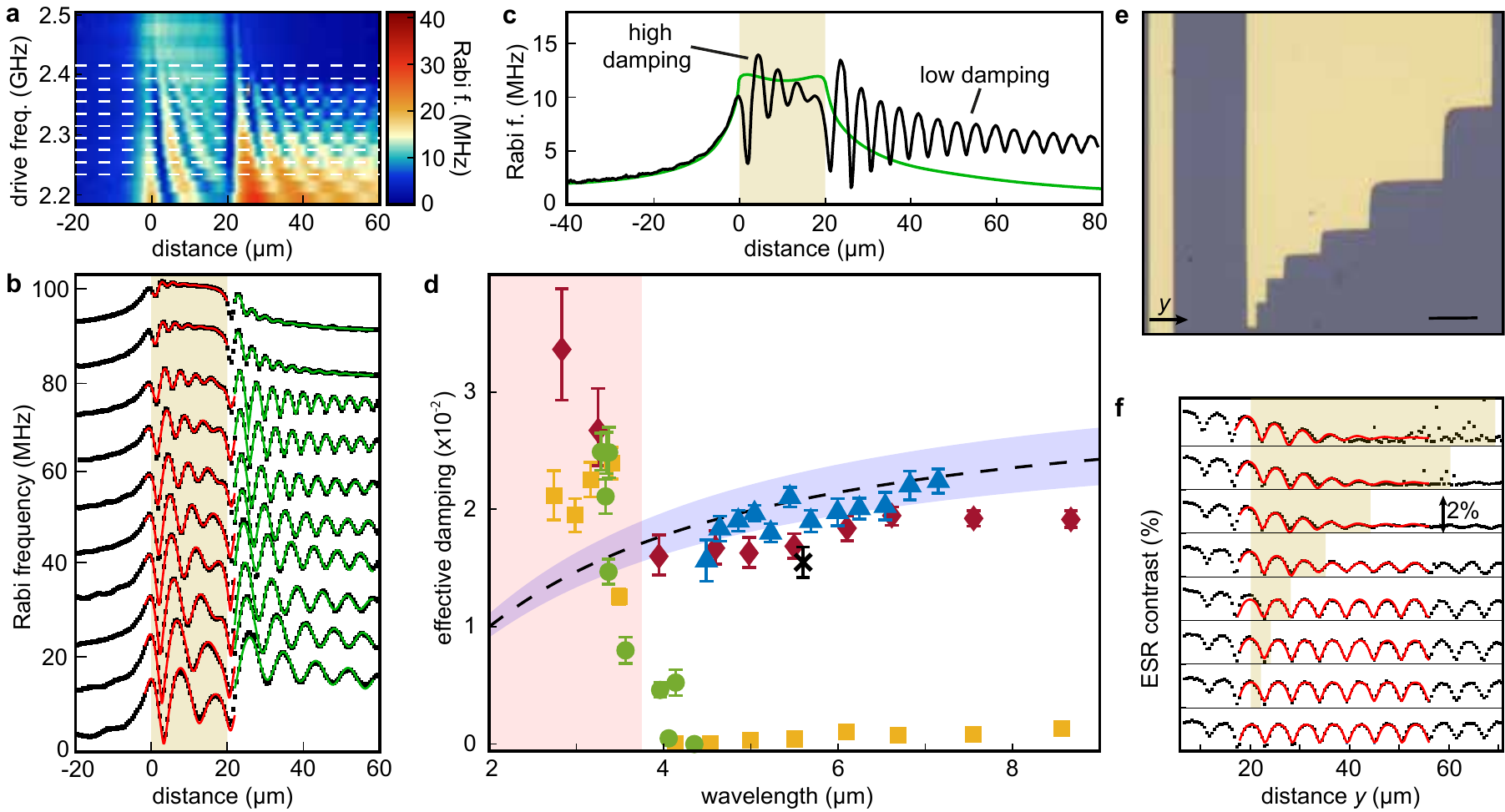}}
\end{center}
\caption{\textbf{Fig. 2. Characterizing metal-induced spin-wave damping.} (\textbf{a}) Imaging spin waves underneath and next to a gold microstrip located between 0-20~$\upmu$m vs drive frequency. Color scale: measured NV Rabi frequency. (\textbf{b}) Linecuts along dashed lines in (a). Black squares: data. Dashed red (green) lines: fits in (next to) the microstrip region. Traces offset by 10~MHz for clarity. (\textbf{c}) Measured NV Rabi frequency (black line) and calculated contribution to it from the direct microstrip field (green line) at drive frequency 2.361~GHz, $B_0=$~18.2~mT. (\textbf{d}) Extracted spin-wave damping versus spin-wavelength for different devices. Yellow squares (red diamonds): device in (a-b), next to (under) the microstrip. Black cross: data from Fig.~\ref{fig2-damping}e-f. Blue triangles: wavelength-dependent damping underneath gold structure (Fig.~\ref{fig2-damping}e, Supplementary Fig.~\ref{fig:ESR_C}). Green circles: data from Fig.~\ref{fig3-damping}. Error bars: $\pm1$ s.d. fit uncertainty. Dashed black line: theoretical model (Eq.~\ref{eq:eq1}) with shaded purple area indicating $10\%$ uncertainty in gold resistivity. Shaded red area: three-magnon scattering region. (\textbf{e}) Micrograph of microstrip and gold structure used in (f). Scale bar: 10~$\upmu$m. (\textbf{f}) ESR contrast along sections of varying length of the gold structure in (e). For each trace, the shaded yellow area indicates the gold structure length. The right microstrip edge is located at $y=5$~$\upmu$m. $B=20.3$~mT, drive frequency $\omega/2\pi= 2.302$~GHz, spin-wavelength$=5.6(2)$~$\upmu$. Black squares: data (for clarity, each trace is offset by $2\%$). Red lines: fits to a model that calculates the ESR contrast by summing the microstrip and spin-wave fields (Supplementary Section \ref{sec:DataFitting}).}
\label{fig2-damping}%
\end{figure}

\noindent To characterize the metal-induced spin-wave damping, we start by analyzing the spatial spin-wave profiles underneath and next to a gold microstrip that we use to excite spin waves (Fig.~\ref{fig2-damping}a). We select a section of microstrip that is far away from corners ($>100$~$\upmu$m) to avoid edge effects. We apply a static magnetic field with in-plane component along the microstrip direction and a drive frequency between 100-600~MHz above the ferromagnetic resonance (FMR), resulting in directional spin-wave emission with a large (small) spin-wave amplitude to the right (left) of the microstrip (Fig.~\ref{fig2-damping}a). This directionality is characteristic of microstrip-driven spin waves traveling perpendicularly to the magnetization and is a result of the handedness of the microstrip drive field and the precessional motion of the spins in the magnet\cite{Mohseni2019,Yu2019a}. We spatially quantify the amplitude of the local microwave magnetic field generated by the spin waves by measuring the rotation rate (Rabi frequency) of the NV spins\cite{Andrich2017}. The spatial oscillations in the measured NV Rabi frequency result from the interference between the microstrip and spin-wave fields\cite{Bertelli2020}. The spin-wavelength is directly visible from the spatial period of these oscillations. We observe a rapid decay of the oscillations underneath the microstrip (Fig.~\ref{fig2-damping}b), even though the microstrip field is approximately constant in this region (Fig.~\ref{fig2-damping}c). We can thus conclude that this decay is caused by the decay of the spin-wave amplitude. In contrast, the decrease of the amplitude away from the microstrip follows the decay of the direct microstrip field (Fig.~\ref{fig2-damping}c).\\

\noindent By fitting the measured spatial decay in- and outside the microstrip region we can extract the additional spin-wave damping caused by the metal (Supplementary Sections \ref{sec:theory} and \ref{sec:DataFitting}). An accurate description of the measured NV Rabi frequencies (Fig.~\ref{fig2-damping}a-c) is only possible if we allow for different damping constants in- and outside the microstrip region (see also Supplementary Figure~\ref{fig:damping}). We find that the damping underneath the gold microstrip (Fig.~\ref{fig2-damping}d, red diamonds) exceeds the damping next to the microstrip (yellow squares) by approximately two orders of magnitude . \\


\noindent We argue that the observed strong spin-wave damping underneath the metal is caused by eddy currents that are induced by the oscillating magnetic stray field of the spin waves. Eddy currents have been reported to cause linewidth broadening of ferromagnetic resonances in cavity and stripline-based experiments\cite{Pincus1960,Kostylev2009,Schoen2015,Li2016,Kostylev2016,Rao2017,Bunyaev2020}.  However, revealing their effect on propagating spin waves, which is important for information transport, has remained an outstanding challenge. We model the effect of the spin-wave-induced currents by including their magnetic field self-consistently into the LLG equation (Supplementary Sections \ref{sec:eddys} and \ref{LLG_eddys}). Doing so, we find that a metal film of thickness $t$ increases the damping to $\alpha= \alpha_\mathrm{G}+\alpha_\mathrm{m}$, with $\alpha_\mathrm{G}$ the intrinsic "Gilbert" damping and
\be
\label{eq:eq1}
\alpha_\mathrm{m}=\gamma\frac{\mu_0^2M_sg^2(k)th}{4\rho}\frac{(1+\eta)^2}{1+\eta^2}
\ee
with $\gamma$ the electron gyromagnetic ratio, $\mu_0$ the vacuum permeability, $M_s$ and $h$ the YIG saturation magnetization and thickness, respectively, $k$ the spin-wavenumber, $\rho$ the metal resistivity, and $\eta$ the spin-wave ellipticity. This expression is derived under the assumption of a homogeneous magnetization across the film thickness $t$, which becomes strictly valid in the thin-film limit $kt\ll 1$. The form factor $g(k)=(1-e^{-kh})(1-e^{-kt})/(k^2 th)\approx 1-k(t+h)$ arises from spatially averaging the dipolar and eddy-current stray fields over the thicknesses of the YIG and metal films. An analysis equating the magnetic energy losses to the power dissipated in the metal yields the same expression (Supplementary Section \ref{sec:effective_damping}). We plot Eq.~\ref{eq:eq1} and its thin-film limit in Fig.~\ref{fig2-damping}d using $\rho =2.44\cdot10^{-8}$~$\Omega$m for the resistivity of gold\cite{Cutnell1997}, finding a good agreement with the damping extracted from the various sets of data without free parameters. The finite width $w$ of the stripline can be disregarded when $kw\gg 1$ (Supplementary Sections \ref{sec:finiteWidth} and \ref{sec:effective_damping}), as is the case in Fig.~\ref{fig2-damping}d. Accounting for a non-homogeneous magnetization may be achieved via micromagnetic simulations\cite{Mohseni2019}\\

\noindent To corroborate the origin of the damping enhancement, we image spin waves propagating underneath a 200-nm-thick gold island deposited next to a microstrip (Fig.~\ref{fig2-damping}e-f). We observe a progressively decreasing spin-wave amplitude for increasing travel distance under the gold, with an average characteristic decay length of $y_0 = 9(1)$~$\upmu$m extracted by fitting the top three traces in Fig.~\ref{fig2-damping}f. We characterize the wavelength dependence by varying the drive frequency (Supplementary Fig.~\ref{fig:ESR_C}). The corresponding damping values are reported in Fig.~\ref{fig2-damping}d (black cross and blue triangles) and agree well with Eq.~\ref{eq:eq1}.\\

\begin{figure}[h]
\begin{center}
{\includegraphics[width=0.5\textwidth]{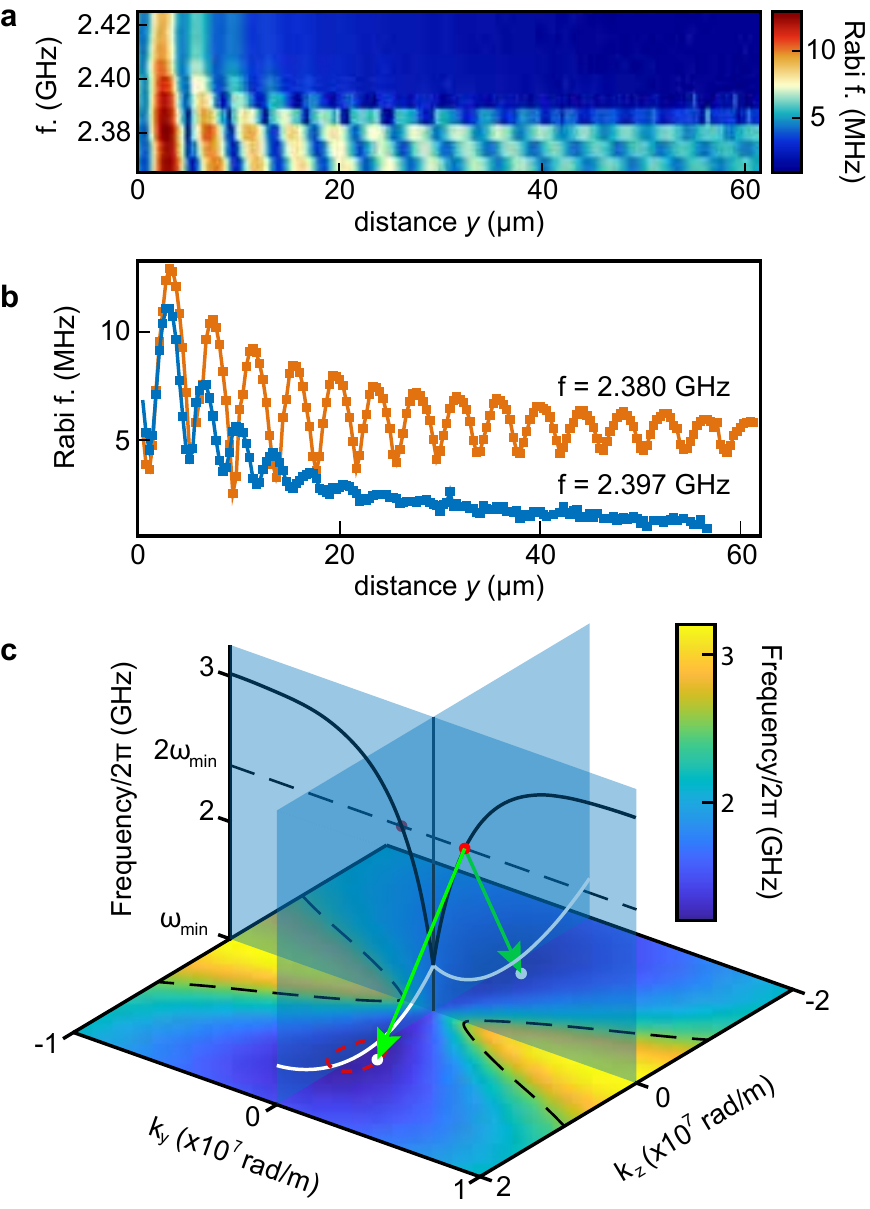}}
\end{center}
\caption{\textbf{Fig. 3. Spatial decay of propagating spin waves caused by three-magnon scattering.} (\textbf{a}) NV Rabi frequency vs spin-wave drive frequency and distance from the edge of the microstrip used for spin-wave excitation. Above a threshold frequency $\omega_\mathrm{T}/2\pi\sim 2.39$~GHz, the spin-wave damping increases strongly. (\textbf{b}) Linecuts of (a) below and above the threshold frequency. When the driving is below the threshold frequency (orange squares and curve), the decrease of the oscillation amplitude follows the decrease of the microstrip field. Above the threshold frequency (blue squares and curve), the spin-wave propagation distance is strongly reduced. (\textbf{c}) Calculated spin-wave dispersion for our 235~nm YIG film. The solid black and white lines show the dispersion along the $y$ and $z$ directions, respectively. The microwave drive excites spin waves propagating along $y$ (red dot, Damon-Eshbach configuration). Above the threshold frequency ($\omega_T=2\omega_min$), scattering of one magnon in this mode into two backward-volume magnons (along $z$, white dots) near the band minimum becomes allowed (red dashed line corresponds to $\sim20$~MHz above $\omega_\mathrm{min}$).}
\label{fig3-damping}%
\end{figure}

\noindent Both in- and outside the stripline region, we observe a sudden increase in damping above a threshold frequency $\omega_\mathrm{T}/2\pi \sim 2.39$~GHz (Fig.~\ref{fig2-damping}a). We characterize this increase in detail by zooming in to the threshold frequency (Fig.~\ref{fig3-damping}a-b) and extracting the damping parameter as a function of the wavelength (Fig.~\ref{fig2-damping}d, green circles). For the spin waves outside the microstrip region, the increase occurs in a $\sim$10~MHz frequency range of the order of the intrinsic spin-wave linewidth.\\

\noindent By analyzing the known spin-wave dispersion of our YIG thin film (Supplementary Section \ref{sec:susceptibility}), it becomes clear that the observed increase in damping above $\omega_\mathrm{T}$ is a result of three-magnon scattering – a process in which one magnon decays into two of half the frequency and opposite wavevectors\cite{Mathieu2003} (Fig.~\ref{fig3-damping}c): When the drive frequency is increased to above $\omega_\mathrm{T}$, three-magnon scattering becomes allowed because $\omega_\mathrm{T}/2$ starts to exceed the bottom of the spin wave band ($\omega_\mathrm{min}$) (Fig.~\ref{fig3-damping}c and Supplementary Fig.~\ref{fig:3ms}). The onset of three-magnon scattering was previously identified using Brillouin light scattering\cite{Schultheiss2009}. Our real-space imaging approach reveals its dramatic effect on the spatial spin-wave decay length important for spin-wave transport. These measurements highlight that damping caused by three-magnon scattering limits the frequency range within which coherent spin waves in YIG thin films can serve as low-damping carriers to $\omega_\mathrm{min}<\omega<2\omega_\mathrm{min}$.\\

\begin{figure}[h]
\begin{center}
{\includegraphics[width=0.5\textwidth]{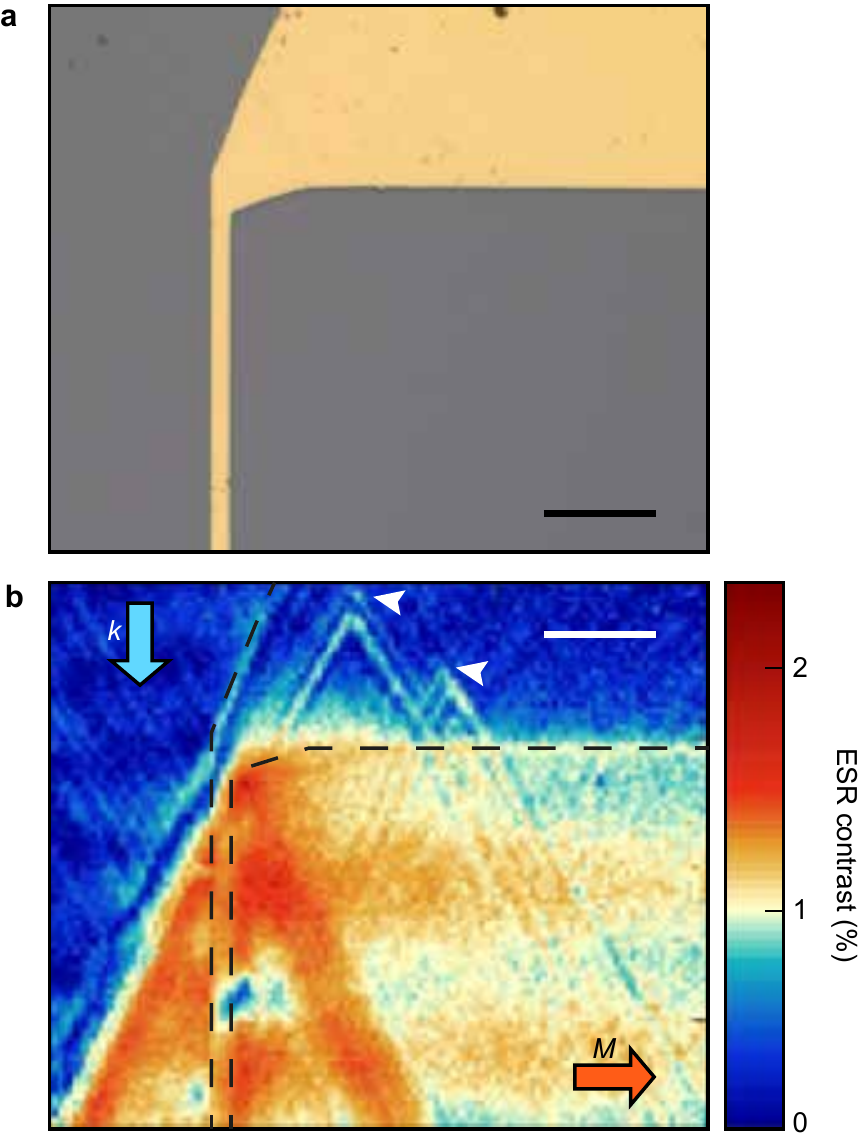}}
\end{center}
\caption{\textbf{Fig. 4. Imaging defect-induced spin-wave scattering underneath a 200~nm metal film.} (\textbf{a}) Micrograph of the gold microstrip used in (b). Scale bar: 20~$\upmu$m. (\textbf{b}) Spatial maps of the NV ESR contrast for $B_0=27.5$~mT and drive frequency $\omega/2\pi= 2.099$~GHz. Two scattering centers (white arrows) are located near the top edge of the image (not clearly identifiable from (a)), as deduced by the scattered caustic beams. Dashed black line: edge of the gold structure. The directions of the magnetization ($\mathbf{M}$) and the predominant wavevector ($\mathbf{k}$) excited by the microstrip are indicated. Scale bar: 20~$\upmu$m.}
\label{fig4-damping}%
\end{figure}

\noindent Finally, we demonstrate that the ability to study spin waves underneath metals also enables the detection of hidden spin-wave scattering centers, highlighting the applicability of this approach for assessing the quality of buried magnetic films in multilayer systems. As an example, we show the scattering patterns produced by defects underneath the metal electrodes used for spin-wave excitation (Fig.~\ref{fig4-damping}a-b). The defects produce characteristic v-shaped patterns, resulting from preferential scattering into the "caustic" directions that are associated with the anisotropic dispersion\cite{Schneider2010}, making the source of these spin-wave beams clearly identifiable. NV-based spin-wave imaging could therefore be used as a diagnostic tool for magnetic quality, even when the material of interest is buried under metallic layers in a heterostructure.\\
\newpage
\subsection*{Discussion}
\noindent In conclusion, we characterized the damping enhancement of spin waves that propagate under metallic electrodes used for spin-wave control, and showed that the increase is well explained by a model that introduces the spin-wave-induced currents into the LLG equation. The ability to detect spin waves underneath metals opens up several exciting new possibilities for studying the interaction between metals and magnets. One example is studying the spectral properties of temperature- or chemical-potential-driven magnon condensates underneath gates in magnon transistors\cite{Cornelissen2018,Wimmer2019,Schneider2020}. Additionally, varying the thickness of the metal and/or magnetic films, or using spacer layers, enables a characterization of interfacial effects such as damping and anti-damping of magnons controlled by the spin-Hall effect in heavy metal electrodes. Furthermore, characterizing the screening of the spin-wave stray fields by a metal enables measuring its magnetic susceptibility at well-defined wavenumbers and extracting material parameters such as skin depth, conductivity and permeability. Finally, the ability to reveal buried scattering centers provides a new tool for assessing the quality of magnetic interfaces and spin-wave devices.\\
\section{Materials and Methods}
\label{sec:Methods}
\subsection*{Sample fabrication}
The diamond chip used in this work measured $2\times 2\times 0.05-$mm$^3$ and had an estimated NV density of $10^3/ ~ \upmu$m$^2$ created via ion implantation at a depth of $\sim$10-20~nm below the diamond surface (see fabrication details in\cite{Bertelli2020}). The YIG film was 235~nm thick, grown on a 500~$\upmu$m-thick GGG substrate via liquid phase epitaxy (Matesy gmbh). The saturation magnetization was previously measured\cite{Bertelli2020} to be $M_s=1.42\cdot10^5$~A/m. To mount the NV-diamond, we deposit a drop of isopropanol onto the YIG and place the diamond on top with the NV-surface facing down, while gently pressing down until the IPA has evaporated. The resulting diamond-YIG distance is limited by small particles (e.g. dust). We extract an NV-YIG distance of 1.6(1)~$\upmu$m from the measured maps of the NV Rabi oscillations.
\subsection*{NV-based imaging of spin waves}
NV centers are optically addressed using a home-built confocal microscope with a 515~nm laser, an NA=0.95 objective for laser focusing/photon collection, and an avalanche photodiode for NV photon detection (for details of the setup, see\cite{Bertelli2020}). The ESR transition of the NV centers used in this work for spin-wave imaging is tuned by a magnetic field $B_0$ according to $\omega = D - \gamma B_0$ where $\gamma/2\pi = 28$~GHz/T is the electron gyromagnetic ratio and $D/2\pi = 2.87$~GHz is the zero-field splitting. In all experiments in Figs. 2-4, the magnetic field is oriented at a $54^\circ $ angle with respect to the sample-plane normal and with an in-plane projection along the microwave stripline, thus aligning it with one of the four possible crystallographic orientations of the NV centers in the diamond. The fields used in this work are below $\sim$25~mT, much smaller than the YIG saturation magnetization ($\mu_0 M_s = 178$~mT), therefore the YIG magnetization tilts out of plane by less than $5^\circ$\cite{Bertelli2020}. We measure Rabi oscillations by applying a $\sim1$~$\upmu$s laser pulse to polarize the NV spin into the $\mathrm{m_s=0}$ state, applying a microwave magnetic field at the NV ESR frequency, and reading out the final spin state through the NV's spin-dependent photoluminescence\cite{Rondin2014}.
\section*{Acknowledgements}
\textbf{Funding:} This work was supported by the Dutch Research Council (NWO) as part of the Frontiers of Nanoscience (NanoFront) program through NWO Projectruimte grant 680.91.115.\\
\textbf{Author contributions:} I.B. and T.S. designed the experiment. B.S. prepared the diamond membrane. I.B. realized the NV magnetometry setup, fabricated the sample and performed the measurements. G.E.W.B., T.S, Y.M.B. and T.Y. developed the theoretical model. J.A. commented on the manuscript. I.B and T.S. analyzed the data and wrote the manuscript with help from all co-authors.\\
\textbf{Data and materials availability:} All data contained in the figures will be made available at 10.5281/zenodo.4726771 upon publication. Additional data related to this paper may be requested from the authors.\\
\textbf{Competing interests:} The authors declare no competing interest.

\newpage
\section{Supplementary Material}
\label{sec:SM}

\subsection{Eddy-current contribution to spin-wave damping}
\label{sec:theory}
In this section we derive the additional spin-wave damping caused by the spin-wave-induced eddy currents in a nearby metallic layer. We use the Landau-Lifshitz-Gilbert (LLG) equation to evaluate the various components of the effective magnetic field and find solutions in the absence of additional damping. Then, we evaluate the spin-wave field inside the metal, derive the eddy currents excited by that field, and calculate the additional field component that acts back on the spin-waves, leading to an expression for the effective damping. Last, we consider the finite width of the metal film in the y direction and include this into the effective damping result.\\
We consider a thin film of a magnetic insulator (i.e. YIG) in the $yz$ plane, between $-t<x<0$, with unit magnetization $\mathbf{m(r)}$ oriented along $z$ in equilibrium and saturation magnetization $M_s$. The bias magnetic field is applied along $z$. The system is translationally invariant along $z$. 

\subsubsection{LLG equation}
\label{sec:LLGs}
The LLG equation is \cite{Gilbert2004}
\be\label{eq:LLG_equations}
\dot{\bfm} = -\gamma \bfm\times \left[\bfB_\text{eff}+\bfB_\text{AC}\right]-\alpha \dot{\bfm}\times \bfm,
\ee	
where $\mathbf{B}_\mathrm{AC}$ is the microstrip magnetic field, $\gamma$ is the gyromagnetic ratio, $\alpha$ is the Gilbert damping and the effective magnetic field is 
\be 
B_{\text{eff},\alpha}= -\frac{1}{M_s}\frac{\partial F}{\partial m_{\alpha}},
\ee
where $\alpha = x,y,z$. 
We will now evaluate the various components of the effective magnetic field. We will assume that the spin-wavelength is much larger than the film thickness ($kt\ll 1$) such that we can approximate the magnetization to be homogeneous across the film thickness. 

The free energy density includes contributions from the external field $\bfB_0$, the demagnetizing field $\mathrm{\bfB_d}$, and the exchange interaction:
\be
F = -M_s\mathbf{m}\cdot(\mathbf{B}_\text{0} + \mathbf{B}_d/2) + \frac{D}{2}
\sum_{\alpha,\beta=x,y,z}\left(\frac{\partial m_{\alpha}}{\partial \beta}\right)^2,
\ee
with $D$ the spin stiffness. We define, for convenience, $ \omega_B = \gamma B_0$, $\omega_M =  \gamma \mu_0 M_s$, and $\omega_{D} = \frac{\gamma D}{M_s}$.  \\
\subsubsection{Evaluating the contributions to the effective magnetic field}
\label{sec:FieldContribution}
\subsubsection*{Zeeman energy}
The Zeeman energy associated with the external magnetic field $\bfB_0 =\omega_B\uz/\gamma$ is 
\be\label{eq:Zeeman Free Energy}
F_z=-M_s\bfm\cdot\bfB_0.
\ee
\subsubsection*{Exchange energy}
The exchange energy density in YIG is isotropic
\be
F_\text{ex}(\bfr) = \frac{D}{2}\sum_{\alpha,\beta=x,y,z}\left(\frac{\partial m_{\alpha}(\bfr)}{\partial \beta}\right)^2.
\ee
Its Fourier transform over the in-plane coordinates $y,z$ is:
\be
F_\text{ex}(\bfk,x) = -k^2 D(m_{y}^2(\bfk,x) + m_{z}^2(\bfk,x)) +  \frac{D}{2}\sum_{\alpha=x,y,z}\left(\frac{\partial m_{\alpha}(\bfk,x)}{\partial x}\right)^2.
\ee
For a constant magnetization over the film thickness, the exchange energy contributes an effective field with Cartesian components: 

\be 
B_{\text{D},\alpha}= -\frac{1}{M_s}\frac{\partial F}{\partial m_{\alpha}} = -\frac{\omega_D}{\gamma}k^2m_{\alpha}(\bfk,x).
\ee

\subsubsection*{Demagnetizing field}

The magnetic field generated by a magnetization $M_s\bfm(\bfr)$ is given by \cite{Guslienko2011}:
\be \label{eq:DemagField}
\bfB(\bfr) = \mu_0M_s \int \Gamma(\bfr-\bfr') \bfm(\bfr') d\bfr',
\ee
where $\Gamma(\mathbf{r-r'})$ is the real-space dipolar tensor, with components that are derivatives of the "Coulomb kernel":
\be
\Gamma_{\alpha\beta}(\bfr) =  \frac{\partial^2}{\partial\alpha\partial\beta}\frac{1}{4\pi|\bfr|},\quad\text{with} \quad\alpha,\beta=x,y,z.
\ee
The 2D Fourier transform of Eq. (\ref{eq:DemagField}) is \footnote{We define $g(k_x) = \int g(x)e^{-i k_xx}dx$ and $g(x) = \frac{1}{2\pi} \int g(k_x)e^{i k_xx}dk_x$}:
\be
\bfB(\bfk,x) = \mu_0M_s \int \Gamma(\bfk,x-x') \bfm(\bfk,x') dx',
\label{eq:Bconv_z}
\ee
where $\bfk=(k_y,k_z)$ and with magnetization 
\be
\bfm(\bfr) = 
\begin{cases}
	\bfm(y,z) & \text{for $-t<x<0$}\\
	0 & \text{elsewhere}
	\end{cases}.
\ee
The demagnetizing field, averaged over the film thickness, is given by:
\be
\overline{\bfB}(\bfk) = \mu_0M_s \frac{1}{t} \int_{-t}^0 \int_{-t}^0 \Gamma(\bfk,x-x') dx'dx \bfm(\bfk) = \mu_0M_s \overline{\Gamma}(\bfk) \bfm(\bfk),
\ee
where the overline indicates averaging over the thickness. The components of the dipolar tensor in Fourier space are:
\be
\Gamma_{\alpha\beta}(\bfk,x) = \frac{1}{2}
\begin{cases}
e^{-k|x|} k  - 2 \delta(x)&\text{for $\alpha = \beta = x,$}\\ 
-e^{-k|x|}\frac{k_\alpha k_\beta}{k} &\text{for $\alpha,\beta = y,z,$}\\
-e^{-k|x|}\text{sign}(x) i k_\alpha &\text{for $\alpha = y,z$ and $\beta=x.$}\\
  
  \label{eq:FFT_Gamma}
\end{cases}
\ee
Using
\begin{align}
&\frac{1}{t}\int_{-t}^0 \int_{-t}^0 e^{-k|x-x'|} dx'dx = \frac{2}{k}(1-\frac{1-e^{-kt}}{kt}) = \frac{2}{k}f(kt), \\
&\frac{1}{t}\int_{-t}^0 \int_{-t}^0 \text{sign}(x-x')e^{-k|x-x'|} dx'dx = 0,\\
& \frac{1}{t}\int_{-t}^0 \int_{-t}^0 \delta(x-x') dx'dx= 1,
\end{align}
we arrive at
\be
\overline{\bfB}(\bfk) = \mu_0 M_s
\begin{pmatrix}\label{eq:Gamma_k}
 f(kt)-1   &   0   &   0 \\
0 & \frac{-k_y^2}{k^2}f(kt) & \frac{-k_yk_z}{k^2}f(kt)   \\	
0 & \frac{-k_yk_z}{k^2}f(kt) & \frac{-k_z^2}{k^2}f(kt)   \\
\end{pmatrix} 
\begin{pmatrix}
m_x(\bfk) \\
m_y(\bfk) \\
m_z(\bfk)	
\end{pmatrix},
\ee
with $f(kt)\rightarrow kt/2$ for $kt\ll 1$.

\subsubsection{Spin-wave susceptibility}
\label{sec:susceptibility}
The linearized Eq. (\ref{eq:LLG_equations}) in the frequency domain reads:
\begin{align}
-i\omega m_{x} &  =-\gamma(B_{z}m_{y}-B_{y})+i\alpha\omega m_{y},%
\label{eq:LLG_x}\\
-i\omega m_{y} &  =-\gamma(B_{x}-B_{z}m_{x})-i\alpha\omega m_{x}.%
\label{eq:LLG_y}%
\end{align}
Using $\mathbf{B}=\mathbf{B}_{\mathrm{eff}}+\mathbf{B}_{\mathrm{AC}}$ and with
$\Gamma_{xy}=\Gamma_{yx}=0$ (from Eq. (\ref{eq:Gamma_k})) we obtain
\begin{align}
\gamma B_{x} &  =\omega_{M}(f-1)m_{x}-\omega_{D}k^{2}m_{x}+\gamma
B_{\mathrm{AC},x},\label{eq:Bx_sol}\\
\gamma B_{y} &  =-\omega_{M}f\sin^{2}\phi m_{y}-\omega_{D}k^{2}m_{y}+\gamma
B_{\mathrm{AC},y},\\
\gamma B_{z} &  =\omega_{B},\label{eq:Bz_sol}%
\end{align}
where $\phi$ is the angle between the wave vector $\mathbf{k}$ and
$\mathbf{B}_{\mathrm{eff}}$. With
\begin{align}
\omega_{0} &  =\omega_{B}+\omega_{D}k^{2} \label{eq:omega0},\\
\omega_{2} &  =\omega_{0}+\omega_{M}(1-f) \label{eq:omega2},\\
\omega_{3} &  =\omega_{0}+\omega_{M}f\sin^{2}\phi \label{eq:omega3},
\end{align}
we obtain Eqns. (\ref{eq:LLG_x}-\ref{eq:LLG_y}) in matrix form:
\begin{equation}%
\begin{pmatrix}
\label{eq:LLG_FM}\omega_{2}-i\alpha\omega & i\omega\\
-i\omega & \omega_{3}-i\alpha\omega
\end{pmatrix}%
\begin{pmatrix}
m_{x}\\
m_{y}%
\end{pmatrix}
=\gamma%
\begin{pmatrix}
B_{AC,x}\\
B_{AC,y}%
\end{pmatrix}.
\end{equation}
Inverting Eq. (\ref{eq:LLG_FM}) gives the susceptibility
\begin{equation}
\chi=\frac{\gamma}{(\omega_{2}-i\alpha\omega)(\omega_{3}-i\alpha\omega
	)-\omega^{2}}%
\begin{pmatrix}
\omega_{3}-i\alpha\omega & -i\omega\\
i\omega & \omega_{2}-i\alpha\omega
\end{pmatrix}
.\label{eq:susc_FM}%
\end{equation}
It is singular when:
\begin{equation}
\label{eq:dispersion}
\Lambda = (\omega_{2}-i\alpha\omega)(\omega_{3}-i\alpha\omega)-\omega^{2}%
=0.
\end{equation}
The real parts of the solutions of this quadratic equation give the spin wave
dispersion $\omega_{sw}=\sqrt{\omega_2\omega_3}$, plotted in Fig.~3c of the main text. In Fig.~3c, the solid lines indicate the dispersion for spin waves propagating along $\pm z$ (i.e., $\phi=0$ and $\phi=\pi$) and along $\pm y$ (i.e., $\phi=\pm\pi/2$). The spin-wave linewidth $\alpha(\omega_2+\omega_3)/2$ follows from the imaginary part of \eref{dispersion}, and the ellipticity of the magnetization precession is given by 
\begin{equation}
\eta=\left|\frac{\chi_{xx}}{\chi_{yx}}\right|_{(\omega=\omega_{sw})}  = \sqrt{\frac{\omega_3}{\omega_2}}.
\end{equation}
Applying the bias field $B_0$ along $\theta_{B_0}=34^\circ$ as in the experiments changes $\omega
_{0}\rightarrow\omega_{B}\cos\theta_{B_{0}}+\omega_{D}k^{2}$, but does not introduce additional terms in the susceptibility for $B_0$ much smaller than the demagnetizing field ($B_0\ll\mu_0 M_s$), as in this work.

\subsubsection{Eddy-current-induced spin-wave damping}
\label{sec:eddys}
In this section, we introduce the field generated by eddy currents into the LLG equation. We first derive the eddy currents in a metal film (parallel to the $yz$ plane and located between $0<x<h$) induced by the spin-wave stray field. The eddy currents in turn generate a magnetic field $\bfB_e$ that couples back into the LLG equation, which should be solved self-consistently. We focus on spin waves travelling in the $+y$-direction, such that $k=k_y$ (thus $\phi=\pi/2$). Our films are much thinner than the magnetic skin depth ($1.7$ $\upmu$m for gold at 2 GHz) such that the dipolar stray fields are not screened significantly. Because the film is thin, we neglect eddy currents in the out-of-plane direction. The in-plane eddy currents are induced by the out-of-plane component of the magnetic field, given by (see \eref{FFT_Gamma}):
\begin{align}\label{eq:B_of_m}
\overline{B}_{x} = &\frac{\mu_0 M_s}{2} \frac{1}{h} \int_0^h dx \int_{-t}^0 dx'k e^{-k(x-x')} (m_{x} - im_{y})  \\  
= &\frac{\mu_0 M_s}{2}kt g (m_{x} - im_{y}),
\label{eq:spinwavefield}
 \end{align}
where the overbar denotes an average over the metal ($h$) thickness. Here, 
\be
g = \frac{(1-e^{-kh})}{kh}\frac{(1-e^{-kt})}{kt}.
\ee
For an infinitely thin film, $g\rightarrow 1$. From Faraday's law, $\overline{B}_x$ generates a charge current :
\begin{align}
J_{z}= & \sigma E_{z} = \sigma\frac{\omega}{k_y} \overline{B}_{x} = \omega \frac{\sigma\mu_0 M_st}{2}  g (m_{x} - im_{y}),
\end{align}
where $\sigma$ is the conductivity and $E_{z}$ the electromotive force. As we will further discuss in \ref{sec:finiteWidth}, this equation is valid in the limit $kw\gg 1$, with $w$ the width of the film, since we used a Fourier transform over $y$ and did not specify boundary conditions. In Fig.~2d of the main text, $w=20$ $\upmu$m and $\lambda< 9$ $\upmu$m, such that $kw > 14$.

\paragraph*{Field generated by the eddy currents}
The current $J_{z}$ generates a field $\bfB_e$ inside the YIG film. Its average over the YIG thickness is 
\begin{align}\label{eq:B_eddy}
\overline{B}_{e,x} = &i\frac{\mu_0J_{z} h}{2}  g = i\omega\frac{\mu_0^2M_s\sigma }{4} th\cdot g^2  (m_{x} - im_{y}),\\
\overline{B}_{e,y} = &i\overline{B}_{e,x} ,
\end{align}
which we can rewrite as
\begin{align}
\gamma \overline{B}_{e,x} = & i \omega \alpha_{m} (m_{x} -i m_{y}),\\
\gamma \overline{B}_{e,y} = & -\alpha_{m}\omega(m_{x} -im_{y}),
\end{align}
where 
\be
\alpha_{m} =  \gamma\frac{\mu_0^2M_s\sigma }{4}th\cdot g^2  
\label{eq:damping}
\ee
is a dimensionless factor that turns out to be the eddy current contribution to the damping as discussed in the next section. Because the equation was derived under the approximation of a homogeneous magnetization across the film thickness it is valid in the thin-film limit $kt,kh\ll1$ where $g^2(k)\rightarrow 1-k(t+h)$. The factor $g^2(k)$ arises from averaging the dipolar and eddy current stray fields over the thicknesses of the metal and magnet films. Including a non-homogeneous magnetization across the film thickness may be achieved via micromagnetic simulations. In Fig.~2d of the main text, $0.16<kt<0.37$ (for $4$ $\upmu$m $<\lambda< 9$ $\upmu$m.)

\subsubsection{Solutions to the LLG equations with eddy currents}
\label{LLG_eddys} We now incorporate $\mathbf{B}_{e}$ into the LLG equation by
adding it to Eqs.\ (\ref{eq:Bx_sol}-\ref{eq:Bz_sol}) for $\phi=\pi/2$ 
\begin{align}
\gamma B_{x} &  =-(\omega_{M}(1-f)+\omega_{D}k^{2})m_{x}+\alpha_m\omega\left(
im_{x}+m_{y}\right)  +\gamma B_{AC,x},\\
\gamma B_{y} &  =-(\omega_{M}f+\omega_{D}k^{2})m_{y}-\alpha_m\omega\left(
m_{x}-im_{y}\right)  +\gamma B_{AC,y},\\
\gamma B_{z} &  =\omega_{B}.\label{eq:field_inplane}%
\end{align}
The linearized LLG equations (\ref{eq:LLG_x}-\ref{eq:LLG_y}) become
\begin{align}
-i\omega m_{x} &  =-(\omega_{3}-i(\alpha+\alpha_m)\omega)m_{y}-\alpha_m\omega m_{x}+\gamma
B_{AC,y},\\
-i\omega m_{y} &  =(\omega_{2}-i(\alpha+\alpha_m)\omega)m_{x}-\alpha_m\omega m_{y}-\gamma
B_{AC,x},
\end{align}
where $\omega_{2}$ and $\omega_{3}$ are given in Eqs.\ (\ref{eq:omega0}-\ref{eq:omega3}). In matrix form:
\begin{equation}%
\begin{pmatrix}
\omega_{2}-i(\alpha+\alpha_m)\omega & (i-\alpha_m)\omega\\
-(i-\alpha_m)\omega & \omega_{3}-i(\alpha+\alpha_m)\omega
\end{pmatrix}%
\begin{pmatrix}
m_{x}\\
m_{y}%
\end{pmatrix}
=\gamma%
\begin{pmatrix}
B_{AC,x}\\
B_{AC,y}%
\end{pmatrix}.
\label{eq:LLG_matrix_eddy}%
\end{equation}
The resulting susceptibility is singular when 
\be
\Lambda = (\omega_{2}-i(\alpha+\alpha_m)\omega)(\omega_{3}-i(\alpha+\alpha_m)\omega) + (i-\alpha_m)^2\omega^2=0.
\ee
Solving this quadratic equation and disregarding terms of order $\alpha^2$ leads to
\be
\omega = \sqrt{\omega_2\omega_3} -i\left[\alpha_m\sqrt{\omega_2\omega_3} + (\alpha+\alpha_m)\frac{\omega_2+\omega_3}{2} \right].
\ee
We observe that including the eddy currents yields the same spin-wave dispersion $\omega_{sw} = \sqrt{\omega_2\omega_3}$, but renormalizes the linewidth according to
\be
\alpha\frac{\omega_2+\omega_3}{2} \rightarrow \alpha_m\left[\sqrt{\omega_2\omega_3}+\frac{\omega_2+\omega_3}{2}\right],
\ee
where we assumed $\alpha_m\gg\alpha$. The eddy-current-induced damping can thus be included into \eref{LLG_equations} by setting 
\be
\alpha = \alpha_e = \alpha_m\frac{\sqrt{\omega_2\omega_3}+\frac{\omega_2+\omega_3}{2}}{\frac{\omega_2+\omega_3}{2}} = \alpha_m \frac{(1+\eta)^2}{1+\eta^2}.
\label{eq:renormalized_damping}
\ee
Substituting \eref{damping} leads to Eq.\ 1 in the main text. In section \ref{sec:effective_damping} we find the same expression using an alternative derivation.

\subsubsection{Metal film of finite width}
\label{sec:finiteWidth}
We now consider a metal strip with finite width $w$ along $y$. The effective orbital magnetization of the eddy currents induced by the spin-wave field points in the $x$-direction and is determined by the Maxwell-Faraday equation:
\begin{equation}
\frac{\partial^{2}m_{z}^{eff}}{\partial y^{2}}=-i\omega\sigma\bar{B}%
_{x}(y),\ \ \ \mathrm{with}\ \ \ j_z=-\partial m_z^{eff}/\partial y,
\label{eq:orbitalMag}
\end{equation}
where $\bar{B}_{x}(y)$ is the stray field of a spin wave travelling in the $+y$ direction, Fourier transformed over time but not over coordinates. It is given by (c.f. \eref{spinwavefield})
\be
\bar{B}_{x}(y) = \frac{\mu_0 M_s}{2}kt g (m_{x} - im_{y}).
\ee
Introducing the notations $m_{x,y}=m_{x,y}^{(0)}e^{iky}$, the solution of \eref{orbitalMag} is
\begin{equation}
m_{x}^{eff}=\alpha_{k}\left(  e^{iky}-\left[1+iky\right]  \frac{\sin
	\frac{kw}{2}}{\frac{kw}{2}}\right),
\end{equation}
with
\begin{equation}
\alpha_{k}=i\omega\sigma\frac{\mu_{0}M_{s}t}{2k}g_{k}(m_{x}^{0}-im_{y}^{0}).
\end{equation}
The eddy-current field averaged over the magnetic film thickness, cf. Eq.\ (\ref{eq:B_eddy}) is:
\begin{equation}
B_{e\alpha}(y)=\frac{1}{t}\int_{-t}^{0}dxB_{e\alpha}(x,y)=\int\frac{dq}{2\pi
}B_{e\alpha}(q)e^{iqy},%
\end{equation}
where
\begin{equation}
B_{e\alpha}(q)=\frac{\mu_{0}M_{s}}{t}\int_{-t}^{0}dx\int_{0}^{h}dx^{\prime
}\int_{-\infty}^{\infty}dydze^{\prime-iqy}\int_{-w/2}^{w/2}dy^{\prime}%
\Gamma_{\alpha x}(\mathbf{r,r^{\prime}})m_{x}^{eff}(\mathbf{r^{\prime}%
})\label{eq:B_ea(q)}.%
\end{equation}
Note that it does not depend on $z$. Using
\begin{equation}
\int_{-\infty}^{\infty}dy^{\prime}dz^{\prime}\frac{e^{iky^{\prime}}}%
{\sqrt{(x-x^{\prime})^2+y^{\prime2}+z^{\prime2}}}=\frac{2\pi}{|k|}%
e^{-|k(x-x^{\prime})|},%
\end{equation}
from Eq. (\ref{eq:B_ea(q)}) we obtain
\begin{align}
B_{e\alpha}(q)  & =\mu_{0}M_{s}\alpha_{k}|q|hg_{|q|}\left\{
\frac{\sin[(k-q)w/2]}{k-q}-2\frac{\sin(qw/2)\sin(kw/2)}{qwk}+\right.  \\
& \left.  -\frac{2}{wq^{2}}\sin\frac{kw}{2}\left[  \sin\frac{qw}{2}-\frac
{qw}{2}\cos\frac{qw}{2}\right]  \right\}
\end{align}
and $B_{e,y}(q)=(iq/|q|)B_{e,x}(q)$. In the wide-strip limit $\lim
_{w\rightarrow\infty}k^{-1}\sin(kw/2)\rightarrow\pi\delta(k)$ such that, back in the real-space and time domains,
\begin{equation}
B_{e\alpha}(y,\tau)=\frac{i\omega\sigma\mu_{0}^2M_{s}}{4}e^{-i\omega \tau}%
th(m_{x}^{0}-im_{y}^{0})(g_{k}^{2}e^{iky}-\frac{2\pi}{w}\delta(k)).
\end{equation}
The last term reflects that a spatially homogeneous mode does not induce eddy currents. The
finite width can be neglected when $kw\gg 1$, in which case we get the same result as Eq.\ (\ref{eq:damping}). In Fig.\ 2d of the main text, $kw > 14$. 

\subsubsection{Effective magnetic damping}
\label{sec:effective_damping}
The effective damping parameter can be derived alternatively by equating the
magnetic and external energy losses \cite{Brataas2011}. According to the LLG
equation the power density per area of a dynamic magnetization for a scalar
Gilbert damping constant reads 
\begin{equation}
p^{\left( m\right) }(y)=-\int \left( \mathbf{\dot{M}}\cdot \mathbf{B}_{%
	\mathrm{eff}}\right) dx=-\frac{\alpha _{\mathrm{G}}M_{s}}{\gamma }\int 
\mathbf{\dot{m}}^{2}dx,
\end{equation}%
where the integral is over the magnetic film thickness. In our geometry the
power loss density of a spin wave mode $\mathbf{m}_{i}$ with index $i$ that
solves the linearized LLG with frequency $\omega _{i}$ is then 
\begin{equation}
p_{i}^{\left( m\right) }(y)=\frac{\alpha _{\mathrm{G}}M_{s}}{\gamma }\omega
_{i}^{2}\int [(m_{i}^{(x)})^{2}+(m_{i}^{(y)})^{2}]dx,
\end{equation}%
In the limit $kt\ll 1$, we can replace $i$ by the wave number $k$ of the
spin wave in the $y$ direction. The time ($\tau $)-dependent magnetization 
\begin{equation}
\mathbf{m}_{k}=m_{k}%
\begin{pmatrix}
\eta _{k}\cos (ky-\omega \tau ) \\ 
\sin (ky-\omega \tau )%
\end{pmatrix}%
\end{equation}%
leads to the time-averaged dissipation 
\begin{equation}
p_{i}^{\left( m\right) }(y)=\frac{\alpha _{\mathrm{G}}M_{s}}{\gamma }\omega
_{k}^{2}\frac{1+\eta _{k}^{2}}{2}\int m_{k}^{2}\left( x,y\right) dx,
\end{equation}%
We model the energy loss per unit of length under the strip by a
phenomenological damping parameter $\alpha _{k}^{\prime }$ as 
\begin{equation}
P_{k}^{\left( m\right) }=tw\frac{\alpha _{k}^{\prime }M_{s}}{\gamma }\omega
_{k}^{2}\frac{1+\eta _{k}^{2}}{2}\overline{m_{k}^{2}},  \label{Pm}
\end{equation}%
where the over-bar indicates the spatial average over the film thickness $t$%
. Assuming that the magnetic skin depth is much larger than the thickness of
the strip $h$, the stray field averaged over the strip thickness above the
film and $k>0$ reads 
\begin{equation}
\bar{B}_{x}=M_{s}\mu _{0}tk\frac{1+\eta _{k}}{2}\overline{m_{k}}\cos \left(
ky-\omega \tau \right) .
\end{equation}%
This field generates an electromotive force (emf) $E_{z}$ according to $%
\partial _{y}E_{z}=-\partial _{\tau }\bar{B}_{x}$: 
\begin{align}
E_{z}(y)& =-\int_{0}^{y}\partial _{\tau }\bar{B}_{x}dy^{\prime }  \label{emf}
\\
& =M_{s}\mu _{0}t\frac{1+\eta _{k}}{2}\overline{m_{k}}\omega \left[ \cos
(ky-\omega \tau )-cos(\omega \tau )\right] +\mathrm{C}.
\end{align}%
The emf does not drive a net charge current since the metal strip is part of
a high impedance circuit.%
\begin{equation}
J_{z}=\sigma t\int_{0}^{w}E_{z}(y)=0
\end{equation}
then fixes the integration constant $\mathrm{C.}$ The time-averaged ($%
\left\langle \cdots \right\rangle $) integrated Ohmic loss per unit length
of the wire then reads 
\begin{align}
P_{k}^{\left( \Omega \right) }& =h\sigma \int_{0}^{w}\left\langle \left\vert
E_{z}\right\vert ^{2}\right\rangle dy \\
& =\sigma \left( \mu _{0}t\overline{m_{k}}\omega \right) ^{2}hw\left( \frac{%
	2\cos kw+k^{2}w^{2}-2}{2\left( kw\right) ^{2}}\right) .  \label{POhm}
\end{align}%
We can now determine the effective damping by setting $P_{k}^{\left( \Omega
	\right) }\equiv P_{k}^{\left( m\right) }.$%
\begin{equation}
\alpha _{k}^{\prime }=\gamma M_{s}ht\sigma \mu _{0}^{2}\frac{\overline{m_{k}}%
	^{2}}{\overline{m_{k}^{2}}}\frac{2\cos kw+k^{2}w^{2}-2}{2\left( kw\right)
	^{2}}\frac{(1+\eta _{k})^{2}}{2(1+\eta _{k}^{2})}.
\end{equation}
In the long-wavelength and wide-metal-strip regime $w^{-1}\ll k\ll t^{-1}$, $%
\overline{m_{k}}^{2}\approx \overline{m_{k}^{2}}$ and $\eta _{k}\approx \eta
,$ 
\begin{equation}
\alpha _{e}^{\prime }=\gamma M_{s}ht\sigma \mu _{0}^{2}\frac{(1+\eta )^{2}}{%
	4(1+\eta ^{2})}
\end{equation}
agrees with Eq.~(\ref{eq:renormalized_damping}). We note that the scalar $%
\alpha _{m}^{\prime }$ should be interpreted as an appropriate average over
the Gilbert damping tensor elements that can be in principle determined by
the same procedure.

\subsection{Data fitting procedures}
\label{sec:DataFitting}
\subsubsection{Extracting the damping from the measured Rabi frequency traces}
\label{sec:ExtractRabi}
To fit the measured Rabi frequencies (Fig.~2a-c of the main text) and extract the spin-wave damping, we follow the procedure described in \cite{Bertelli2020}. In this procedure, we first calculate the magnetic field generated by a microwave current in a microstrip propagating along $z$, given by $\mathbf{B_{AC}} = (B_{AC,x}\ux+B_{{AC},y}\uy)$. We then calculate the resulting magnetization dynamics in Fourier space using $\bfm(\bfk)=\chi(\bfk)\bfB_\text{AC}(\bfk)$. From $\bfm(\bfk)$, we calculate the stray field of the spin waves at the location of the NV sensing layer. We then sum (vectorially) the spin-wave and microstrip fields and calculate the resulting NV Rabi frequency. Free fitting parameters are the microwave current through the microstrip, the spin-wave damping, and a $\sim$ 1~MHz spatially homogeneous offset to account for the field generated by the leads delivering the current to the stripline.
\begin{figure}[htbp]
\centering
\includegraphics[width=0.83\textwidth]{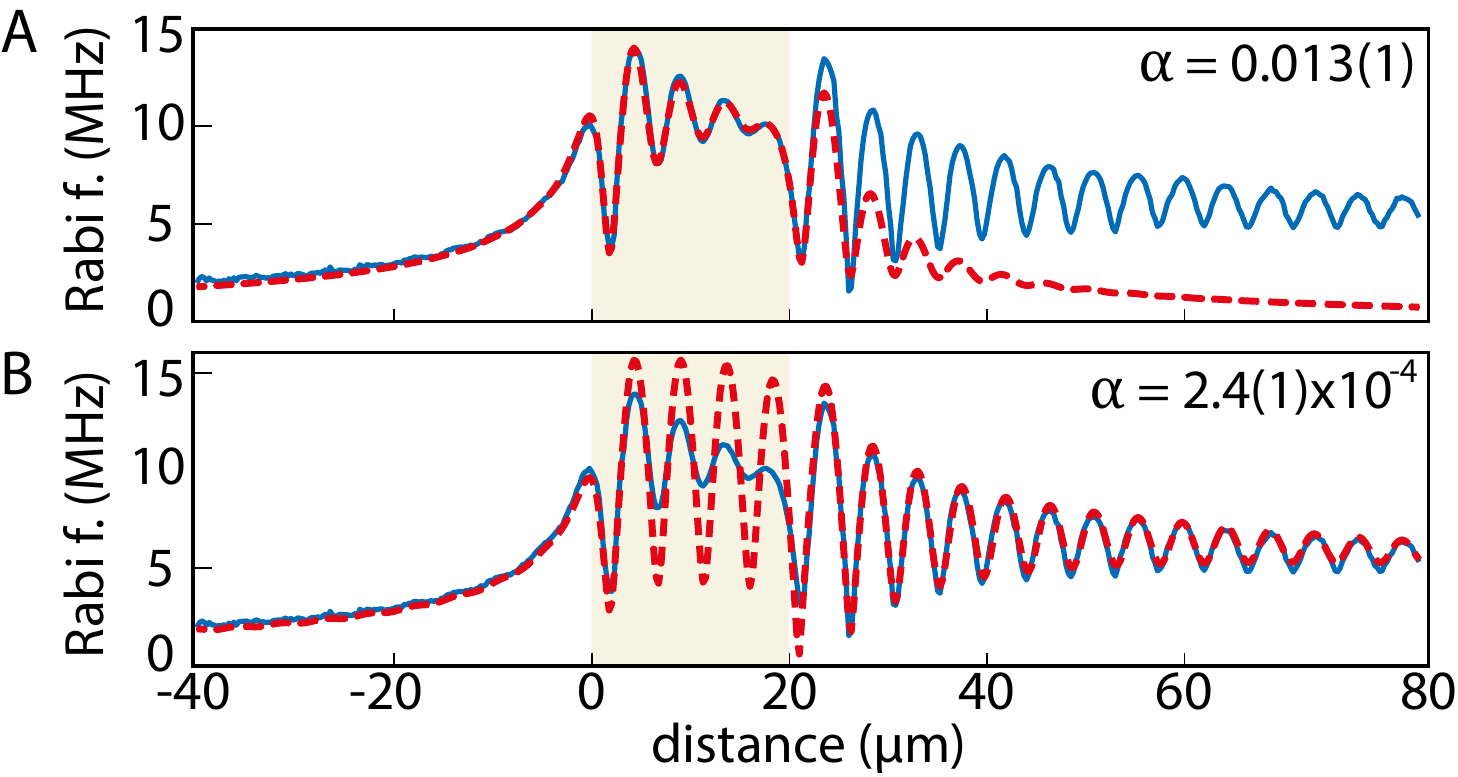}
\caption{\textbf{Highlighting the different spin-wave damping underneath and next to the microstrip.} Solid blue lines: data trace from Fig.\ 2c of the main text. Dashed red lines: calculated Rabi frequencies for high (A) and low (B) values of the damping. The calculations use a single value of the damping for the entire spatial range. The high-damping calculation (panel A) only matches the data well in the microstrip region. The low-damping calculation (panel B) only matches the data next to the microstrip. The microstrip is indicated by shaded yellow color.}
	\label{fig:damping}
\end{figure}\\
Figure \ref{fig:damping} shows two example traces calculated using this procedure (red dashed lines) and compares these to a measured trace (blue line) of Fig.\ 2c of the main text. In both A and B, the calculated traces use a single value of the damping for the entire spatial range. These plots highlight that the measured data in the microstrip region are only described well for a large value of the damping, while the data next to the microstrip are only described well for a low value of the damping.
\subsubsection{Extracting the damping under the gold structure}
\label{sec:ExtractContr}
\begin{figure}[htbp]
\centering
	\includegraphics[width=0.8\textwidth]{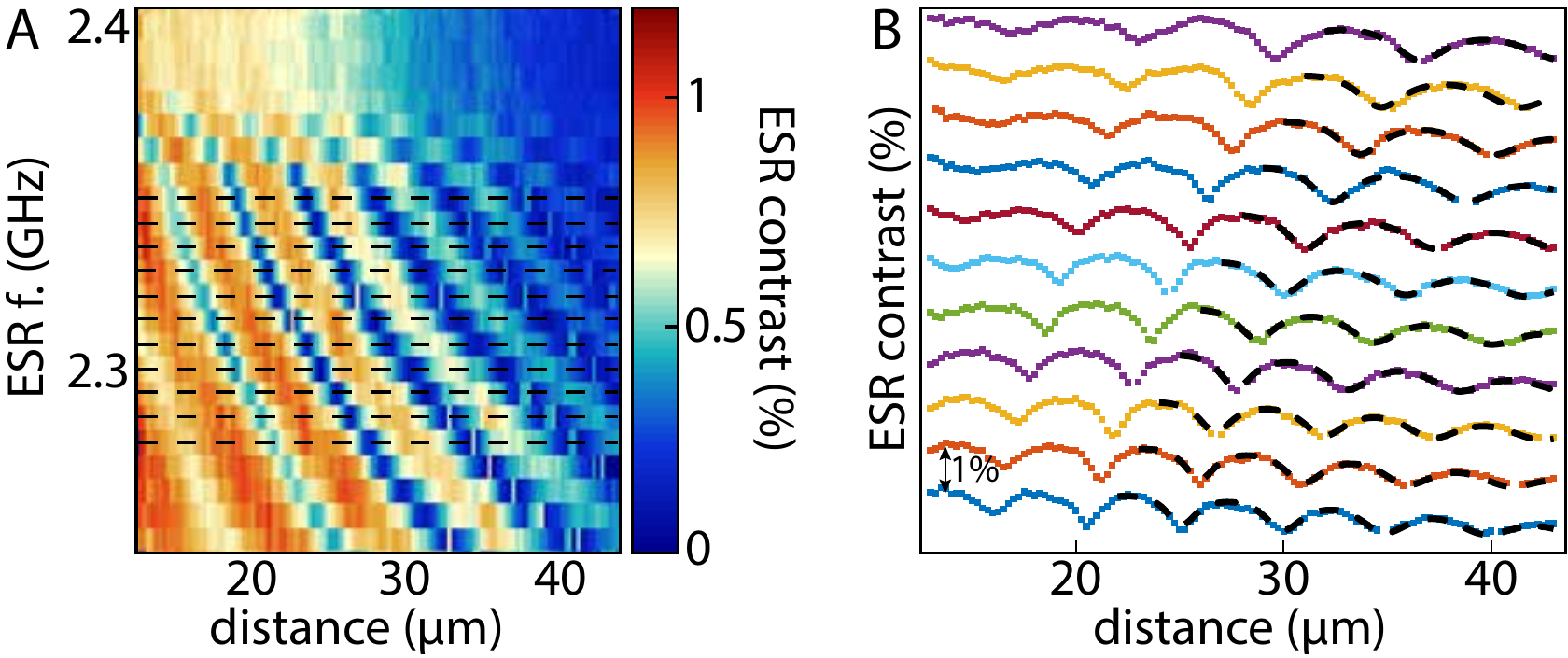}
	\caption{\textbf{Spin-wave damping under gold structure.} (\textbf{A}) ESR contrast vs distance for different spin-wave frequencies under the gold structure shown in Fig.~2e of the main text. Dashed black lines: linecuts shown in (B). (\textbf{B}) Colored lines: linecuts of (A). Dashed black lines: fits. The fitting range was chosen such that it starts at the first peak for which a decay is visible.}
	\label{fig:ESR_C}
\end{figure}

To extract the spatial decay length of the spin waves $y_{decay}$ underneath the gold structure from spatial measurements of the ESR contrast $C(y)$ (Fig.~2e-f of the main text and Supplementary Fig.~\ref{fig:ESR_C}) we describe $C(y)$ using
\begin{equation}
C(y) = C_0\frac{\Omega^2(y)}{\Omega^2(y)+1},
\end{equation}
where $C_0$ is the known maximum ESR contrast and $\Omega(y)$ is a normalized NV Rabi frequency resulting from the sum of the spin-wave and direct microstrip fields:
\begin{equation}
\Omega(y)=\left\vert iAe^{i(k(y-y_0)}e^{-(y-y_\text{struct})/y_{decay}} + \frac{B}{y-y_{0}}\right\vert.
\end{equation}
Here, $y_0$ and $y_\text{struct}$ are the known locations of the edges of the microstrip and gold structure, respectively (see Fig.~2e of the main text), and $A$, $B$, and $y_{decay}$ are extracted from the fits. The spatial decay length $y_{decay}$ is given by the linewidth of the susceptibility in $k$-space and can therefore be related to the damping parameter $\alpha$ by Taylor expanding $\omega_{sw}(k)\approx \omega_{sw}(k_0)+v_g(k-k_0)$ in \eref{dispersion} to get:
\be
\Lambda = 2\omega_{sw}\left(v_g(k-k_0) - i\alpha\frac{\omega_2+\omega_3}{2}\right).
\label{eq:Lambda2}
\ee
Solving $\Lambda=0$, we find
\be
k=k_0+i\alpha\frac{\omega_2+\omega_3}{2v_g},
\ee
which yields the relation between the spatial decay length and $\alpha$
\be
y_{decay} = \frac{2 v_g}{\alpha (\omega_2+\omega_3)},
\label{eq:decaylength}
\ee
where we calculate $\omega_2$ and $\omega_3$ (defined in Eqs.\ (\ref{eq:omega2}) and (\ref{eq:omega3})) and the spin-wave group velocity $v_g$ from the spin-wave dispersion.\\
This fit procedure is used to extract the damping from the data in Fig.~2f of the main text, as well as to determine the frequency dependence of the damping underneath the gold structure, for which the data traces and fits are shown in Fig.~\ref{fig:ESR_C}. The extracted values of the damping are plotted in Fig.\ 2d of the main text.

\subsubsection{Three-magnon scattering threshold}
\label{sec:3ms}
\begin{figure}[htbp]
\centering
	\includegraphics[width=0.5\textwidth]{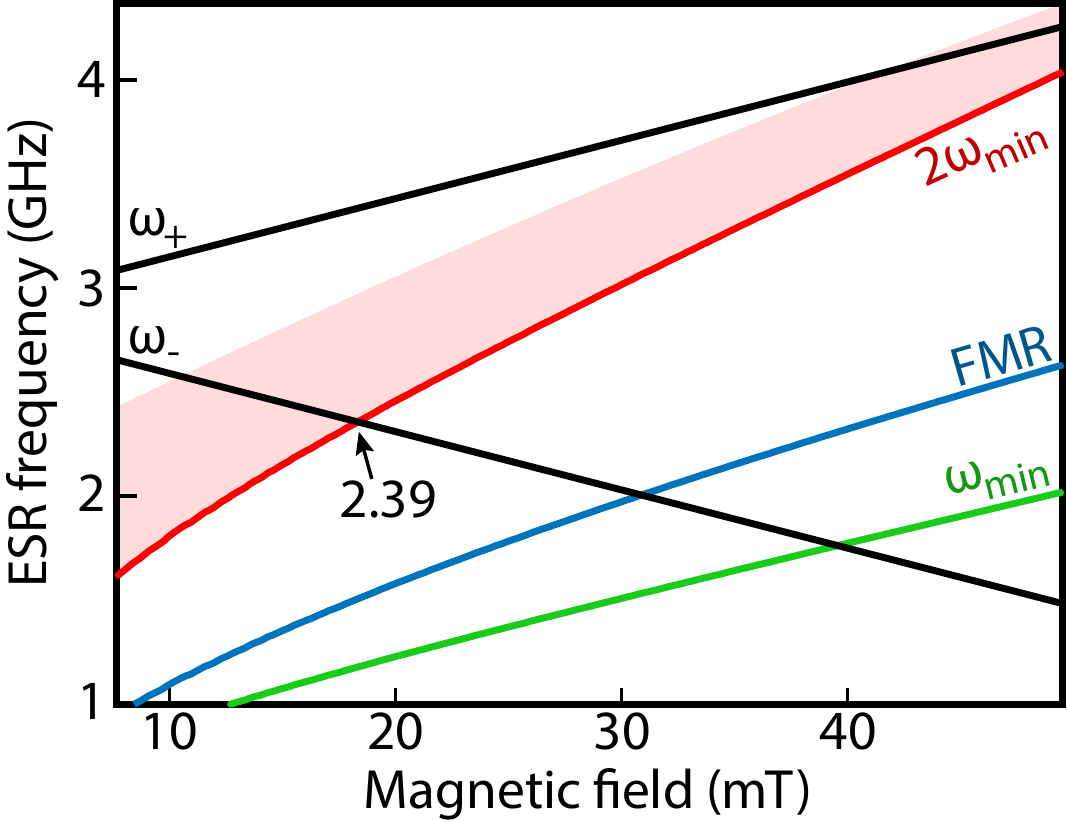}
	\caption{\textbf{Calculated three-magnon scattering threshold frequency vs magnetic field applied along the NV axis.} For frequencies above $2\omega_{\mathrm{min}}$ (shaded red area), scattering of one DE spin wave into two BV waves of frequency close to the band minimum ($\omega_{\mathrm{min}}$, solid green line) becomes possible. Solid black lines (indicated as $\omega_\pm$) represent the NV ESR transitions.  $\omega_-$ and $2\omega_{\mathrm{min}}$ cross at a frequency close to 2.39 GHz, as shown in Fig.~3 of the main text. Solid blue line: FMR of YIG.}
	\label{fig:3ms}
\end{figure}

The three-magnon scattering process is enabled for spin waves of frequency at least twice that of the bottom of the spin-wave band ($\omega_{\mathrm{min}}$), which shifts with the applied magnetic field. In the main text, we see this threshold at $\sim$2.39~GHz (Fig.~3). From the spin-wave dispersion (\eref{dispersion}), we find that this frequency corresponds to the frequency at which the $\omega_-$ NV ESR transition and $2\omega_{\mathrm{min}}$ cross (Fig.~\ref{fig:3ms}).
\newpage
\bibliographystyle{MyNiceStyle4}
\bibliography{BibSWdamping}
\end{document}